\documentclass[a4paper,aps,11pt,nofootinbib,pdftex]{revtex4}
\usepackage{amsmath}
\usepackage{amssymb}
\usepackage{amsfonts}
\usepackage{mathrsfs}
\usepackage{color}
\usepackage{float}
\usepackage{comment}
\usepackage{ulem}
\usepackage{ascmac}
\usepackage{mathtools}
\usepackage{booktabs}
\usepackage{multirow}
\usepackage{graphicx}

\definecolor{DarkBlue}{rgb}{0,0,0.7}

\definecolor{DarkRed}{rgb}{0.65,0,0}

\definecolor{DarkGreen}{rgb}{0,0.3,0}

\definecolor{purple}{rgb}{0.7,0,0.7}

\def\figwidth{5cm}

\def\S{\mathscr{S}}

\begin{document}

\def\uwave{ }
\title{\large
Gravastars as Nontopological Solitons
}

\hfill{TU-1242}

\hfill{OCU-PHYS 600}

\hfill{~AP-GR 198}

\hfill{NITEP 222}


\author{\vspace{1cm}\large Tatsuya Ogawa}
\email{tatsuya.ogawa.d8@tohoku.ac.jp}
\affiliation{
Department of Physics, Tohoku University, Sendai, Miyagi 980-8578, Japan, \\
Osaka Central Advanced Mathematical Institute, Osaka Metropolitan University,
Osaka 558-8585, Japan}
\author{\large Hideki Ishihara}
\email{h.ishihara@omu.ac.jp}
\affiliation{
Nambu Yoichiro Institute of Theoretical and Experimental Physics (NITEP),
Osaka Central Advanced Mathematical Institute, Osaka Metropolitan University,
Osaka 558-8585, Japan
}

\begin{abstract}
\vspace{1cm}
We investigate a coupled system consisting of a complex scalar field, a U(1) gauge field,
a complex Higgs scalar field, and Einstein gravity.
We present nontopological soliton star solutions in which the interior geometry is
described by the de Sitter metric, the exterior by the Schwarzschild metric,
and these two regions are attached by a spherical surface layer with a finite thickness.
This structure is the same as the so-called gravastars,
so we refer to these solutions as \lq solitonic gravastars\rq .
We demonstrate that compact solitonic gravastars that possessing photon spheres can appear.
Additionally, we study the mass and compactness of solitonic gravastars for various sets of parameters
that characterize the system.
\end{abstract}

\maketitle

  \section{Introduction}
Black holes gather much attention as compact astrophysical objects at the
centers of galaxies \cite{Ghez:2000ay,EventHorizonTelescope:2019dse}
and as sources of gravitational waves \cite{LIGOScientific:2016aoc}.
By definition of the black hole, any observable phenomena should occur outside the event horizons.
Therefore, various black hole mimickers that allow the same phenomena as the black holes
are also interesting objects to be investigated.

Gravastars are proposed to be formed by quantum effects in gravitational collapse of stars as
a kind of black hole mimickers \cite{Mazur:2001fv, Mazur:2004fk}.
The interior of a gravastar is filled with vacuum energy, while the exterior is true vacuum.
Therefore, the geometry of the interior region is described by the de Sitter metric,
while the exterior by the Schwarzschild metric.
These two regions are attached by a surface layer with a finite thickness.
The gravastar has neither the central singularity nor the event horizon because the radius of the surface layer is larger than the Schwarzschild radius but smaller
than the de Sitter horizon radius.

On the other hand, it is known that compact objects can be produced as solutions
in a class of bosonic field theories with global U(1) invariance,
known as nontopological solitons (NTSs),
where their stability is guaranteed by the conserved U(1) charge
\cite{Friedberg:1976me, Coleman:1985ki}.
In astrophysics, such NTSs are considered as dark matter candidates
\cite{Kusenko:1997si, Kusenko:2001vu, Fujii:2001xp,Enqvist:2001jd, Kusenko:2004yw}.
Furthermore, NTSs coupled with gravity, known as soliton stars, are widely studied \footnote{
Compact objects described by solutions of a free scalar field coupled with gravity are also studied as boson stars \cite{Kaup:1968zz, Ruffini:1969qy, Colpi:1986ye}.
}
\cite{Lee:1986ts, Friedberg:1986tq, Kunz:2021mbm, Lynn:1988rb, Jetzer:1989av, Ishihara:2018rxg, Ishihara:2019gim, Ishihara:2021iag, Forgacs:2020vcy, Forgacs:2020sms}.
(See also \cite{Schunck:2003kk, Liebling:2012fv} for reviews).
Fermion soliton stars, NTSs composed of a scalar field and an ideal Fermi gas, were proposed in \cite{Lee:1986tr,DelGrosso:2023dmv}.
These soliton stars can also be black hole mimickers.

Recently, we found that \textit{solitonic gravastars}, NTS stars with the structure of the gravastar, exist as solutions \cite{Ogawa:2023ive} in a coupled system consisting of a complex scalar field, a U(1) gauge field, a complex Higgs scalar field, and Einstein gravity
If we choose a set of coupling constants that characterizes the field theory, the solitonic gravastar solutions can be compact enough to have photon spheres.
In this paper, we numerically obtain solutions for solitonic gravastars and investigate their properties, which depend on the coupling constants.

The paper is organized as follows.
In Sec. II, we present the U(1) gauge Higgs model studied in this paper, and derive the basic equations under the assumptions of symmetries of the fields.
In Sec. III, we obtain solitonic gravastar solutions numerically, and investigate their physical properties.
In Sec. IV, we examine how the mass and compactness depend on the model parameters that characterize the system.
Finally, Sec. V is devoted to the summary.

  \section{U(1) Gauge Higgs Model}
We consider the model of field theory described by the action:
\begin{align}
	S=\int \sqrt{-g}d^4x &\left(	\frac{R}{16\pi G}\right.
	-g^{\mu\nu}(D_{\mu}\psi)^{\ast}(D_{\nu}\psi)
	-g^{\mu\nu}(D_{\mu}\phi)^{\ast}(D_{\nu}\phi)
\cr	&
 	-\frac{\lambda}{4}(\left|\phi \right|^2-\eta^2)^2-\mu \left|\psi \right|^2\left|\phi \right|^2
\cr
	&\left.	-\frac{1}{4}g^{\mu\nu}g^{\alpha\beta}F_{\mu\alpha}F_{\nu\beta}\right),
\label{eq:action}
\end{align}
where $R$ is the Ricci scalar of a metric $g_{\mu\nu}$, $g:=\det(g_{\mu\nu})$,
and $G$ is the gravitational constant.
$\psi$ and $\phi$ are complex scalar fields,
$F_{\mu\nu}:=\partial_{\mu}A_{\nu}-\partial_{\nu}A_{\mu}$ is the field strength
of a U(1) gauge field $A_{\mu}$, and
$D_{\mu}:=\partial_{\mu}-ieA_{\mu}$ is the gauge-covariant derivative operator
with a coupling constant $e$.
The spotaneously symmetry breaking occurs by the self-coupling of $\phi$
with a coupling constant  $\lambda$, and a symmetry breaking scale $\eta$.
The parameter $\mu$ is a coupling constant between $\psi$ and $\phi$.

Variation of the action \eqref{eq:action} yields the set of field equations,
\begin{align}
 &\frac{1}{\sqrt{-g}}D_{\mu}\left(\sqrt{-g}g^{\mu\nu}D_{\nu}\psi\right)
 	-\mu \psi \left|\phi \right|^2=0,
\label{eq:ELequation_psi}  \\
 &\frac{1}{\sqrt{-g}}D_{\mu}\left(\sqrt{-g}g^{\mu\nu}D_{\nu}\phi\right)
	-\frac{\lambda}{2}\phi(\left|\phi \right|^2-\eta^2)-\mu \left|\psi \right|^2 \phi=0,
\label{eq:ELequation_phi} \\
 &\frac{1}{\sqrt{-g}}\partial_{\mu}(\sqrt{-g}F^{\mu\nu})=j^{\nu}_{\psi}+j^{\nu}_{\phi},
\label{eq:ELequation_A} \\
 &G_{\mu\nu}=8\pi GT_{\mu\nu},
\label{eq:Einsteinequation}
\end{align}
where $j^{\mu}_{\psi}$ and $j^{\mu}_{\phi}$ are current densities defiend by
\begin{equation}
\begin{split}
	&j^{\mu}_{\psi}:=ie\left(\psi^{\ast}(D^{\mu}\psi)-(D^{\mu}\psi)^{\ast}\psi\right),
\cr
	&j^{\mu}_{\phi}:=ie\left(\phi^{\ast}(D^{\mu}\phi)-(D^{\mu}\phi)^{\ast}\phi\right),
\label{eq:current_phi_psi}
\end{split}
\end{equation}
respectively.
In the equation \eqref{eq:Einsteinequation}, the Einstein tensor is defined by
$G_{\mu\nu}:=R_{\mu\nu}-\frac{1}{2}Rg_{\mu\nu}$,
and $T_{\mu\nu}$ is the energy momentum tensor defined by
\begin{align}
T_{\mu\nu}
=& 2(D_{\mu}\psi)^{\ast}(D_{\nu}\psi)
-g_{\mu\nu}	(D_{\alpha}\psi)^{\ast}(D^{\alpha}\psi)
\cr
&+2(D_{\mu}\phi)^{\ast}(D_{\nu}\phi)
-g_{\mu\nu}(D_{\alpha}\phi)^{\ast}(D^{\alpha}\phi)
\cr
&-g_{\mu\nu}\left( \frac{\lambda}{4}(|\phi|^2-\eta^2)^2
 + \mu |\psi|^2|\phi|^2 \right)
\cr
&+\left( F_{\mu\alpha}F_{\nu}^{~\alpha}
-\frac{1}{4}g_{\mu\nu}F_{\alpha\beta}F^{\alpha\beta}\right).
\label{eq:T_munu}
\end{align}

The action \eqref{eq:action} has the symmetry under the
$\text{U}(1)_\text{local}\times \text{U}(1)_\text{global}$ gauge-transformation,
\begin{align}
    &\psi(x) \to \psi'(x)=e^{i(\chi(x)+\beta)}\psi(x),
 \label{eq:psi_tr} \\
  	&\phi(x) \to \phi'(x)=e^{i(\chi(x)+\gamma)}\phi(x),
 \label{eq:phi_tr} \\
	&A_{\mu}(x)\to A_{\mu}'(x)=A_{\mu}(x)+e^{-1}\partial_{\mu}\chi(x),
 \label{eq:A_tr}
\end{align}
where $\chi(x)$ is an arbitrary function, and $\beta$ and $\gamma$ are arbitrary constants.
Owing to the gauge invariance, conservation laws
\begin{align}
	\partial_\mu(\sqrt{-g}~ j^{\mu}_{\psi})=0
	~\mbox{and}~ \partial_\mu (\sqrt{-g}~j^{\mu}_{\phi})=0
\end{align}
are satisfied.
Total charges of $\psi$ and $\phi$ on the $t=const.$ slices defined by
\begin{align}
	Q_{\psi}:=\int d^3x \sqrt{-g}~\rho_{\psi} , \quad\mbox{and}\quad
	Q_{\phi}:=\int d^3x \sqrt{-g}~\rho_{\phi},
\label{eq:charge_phi_psi}
\end{align}
are conserved, respectively,
where $\rho_{\psi}:=j^t_{\psi}$ and $\rho_{\phi}:=j^t_{\phi}$.

We assume the metric is static and spherically symmetric in the form
\begin{align}
  ds^2=&-\sigma(r)^2F(r)dt^2+F(r)^{-1}dr^2
	+r^2\left(d\theta^2+\sin^2\theta d\varphi^2\right),
  \label{eq:metric} \\
  &F(r):=1-\frac{2Gm(r)}{r},
\nonumber
\end{align}
where $\sigma(r)$, the lapse function, and $m(r)$, the mass function, are assumed to be
dependent on the radial coordinate $r$.
For the metric \eqref{eq:metric}, we see $\sqrt{-g}=\sigma r^2\sin \theta$.
We also assume stationality and spherically symmetry
on $\psi,\phi$ and $A_{\mu}$ in the form
\begin{align}
  	\psi=e^{-i\omega t}u(r), \quad \phi=e^{-i\bar\omega t}f(r), \quad A_{\mu}dx^{\mu}=A_t(r)dt,
  \label{eq:matter_ansatz}
\end{align}
where $\omega$ and $\bar\omega$ are constants, and $u(r), f(r),$ and $A_t(r)$ are functions of $r$.

Using the gauge transformation \eqref{eq:psi_tr} - \eqref{eq:A_tr},
we fix the variables as
\begin{align}
	&\psi(t,r) \to e^{i\Omega t}u(r), \quad     \phi(t,r) \to f(r), \quad
\cr
	&A_t(r) \to \alpha(r):= A_t(r)+e^{-1}\bar\omega,
\label{eq:matter_variable}
\end{align}
where $\Omega:=\bar\omega-\omega$.

Hereafter, we normalize dimensional variables by symmetry breaking energy scale $\eta$ as
\begin{equation}
\begin{split}
  &r\to \tilde{r}=\eta r,~~u\to \tilde{u}=u/\eta,~~f\to \tilde{f}=f/\eta,~~
	\alpha\to \tilde{\alpha}=\alpha/\eta,
\\
	&m\to \tilde{m}=G\eta m,~~\Omega\to \tilde{\Omega}=\Omega/\eta.
\end{split}
\label{dimensionless}
\end{equation}
For simplicity, we drop all tildes.
Substituting \eqref{eq:metric} and \eqref{eq:matter_variable} into \eqref{eq:ELequation_psi} - \eqref{eq:ELequation_A},
we reduce the field equations of $\psi$, $\phi$, and $A_{\mu}$ to
\begin{align}
  &u''+\left(\frac{2}{r}+\frac{\sigma'}{\sigma}+\frac{F'}{F}\right)u'
	+F^{-1}\left(\frac{(e\alpha-\Omega)^2u}{\sigma^2F}
	-\mu f^2u\right)=0,
 \label{eq:equation_u}
\\
 &f''+\left(\frac{2}{r}+\frac{\sigma'}{\sigma}+\frac{F'}{F}\right)f'
	+F^{-1}\left(\frac{e^2f\alpha^2}{\sigma^2F}
	-\frac{\lambda}{2}f(f^2- 1)-\mu fu^2\right)=0,
 \label{eq:equation_f}
\\
  &\alpha''+\left(\frac{2}{r}-\frac{\sigma'}{\sigma}\right)\alpha'
	+F^{-1}\left(-2e^2f^2\alpha -2e(e\alpha-\Omega) u^2\right)=0,
\label{eq:equation_alpha}
\end{align}
where prime denotes the derivative with respect to $r$.

By the assumptions \eqref{eq:metric} and \eqref{eq:matter_ansatz},
we solve the Einstein equations
\begin{align}
 G^t_{~t}=8\pi GT^t_{~t}
\end{align}
\label{eq:Einsteinequation2}
and
\begin{align}
 G^r_{~r}-G^t_{~t}=8\pi G(T^r_{~r}-T^t_{~t}).
\label{eq:Einsteinequation3}
\end{align}
The remaing equations are guaranteed by the Bianchi identity.
The equations for $\sigma$ and $m$ are given by
\begin{align}
 & \frac{2m'}{r^2}- 8\pi G\eta^2\biggl(\frac{e^2f^2\alpha^2}{\sigma^2F}+\frac{(e\alpha-\Omega)^2u^2}{\sigma^2F}
 +F( f'^2 +u'^2)+\frac{\lambda}{4}(f^2-1)^2+\mu f^2u^2 +\frac{1}{2\sigma^2}\alpha'^2 \biggr)=0,
 \label{eq:equation_Rtt}
\\
 &\frac{\sigma'F}{r\sigma}
 	-8\pi G\eta^2 \bigg( \frac{e^2f^2\alpha^2}{\sigma^2F}
		+\frac{(e\alpha-\Omega)^2u^2}{\sigma^2F}+F(f'^2 +u'^2)\bigg)=0.
 \label{eq:equation_RttRrr}
\end{align}
The dimensionless parameter $G\eta^2=\left(\eta/M_P\right)^2$ represents
the ratio of the symmetry breaking scale and Plank mass\footnote{In the limit $G\eta^2\to 0$, the gravitational field decouple
to the matter fields.
If we consider U(1) Higgs model in the flat spacetime,
the equations \eqref{eq:equation_u} - \eqref{eq:equation_alpha} reduce to
ones for non-topological solitons discussed in \cite{Ishihara:2018rxg,Ishihara:2019gim,Ishihara:2021iag}.}.

We impose regularity conditions of the fields at the origin,
\begin{align}
	\frac{d\sigma}{dr}=0, \quad m=0, \quad \frac{du}{dr}=0, \quad
	\frac{df}{dr}=0, \quad \frac{d\alpha}{dr}=0, \qquad \text{at $r=0$.}
\label{eq:BC_origin}
\end{align}
In addition, we require the energy density
$\epsilon:=-T^t_{~t}$ is vanishing at spatial infinity.
Then, we impose the conditions,
\begin{align}
	u=0, \quad f=1, \quad \alpha=0, \qquad \text{at $r=\infty$.}
\label{eq:BC_infinity_matter}
\end{align}
The complex Higgs scalar field $\phi$ takes nonvanishing
vacuum expectation value, and therefore, symmetry is spontaneously broken
in the far region.
As a result, the complex scalar field $\psi$ and the gauge field $A_{\mu}$ aquire
masses $m_{\psi}:=\sqrt{\mu}\eta~$ and $~m_{A}:=\sqrt{2}e \eta$, respectively.
From the requirements \eqref{eq:BC_infinity_matter},
the metric approaches to the Schwarzschild metric, then
\begin{align}
	\sigma=1, \quad m=m_{\infty}=\text{\it const.},
\label{eq:BC_infinity_metric}
\end{align}
at spatial infinity.

\section{Solitonic Gravastar Solutions}
In this section, we solve the equations \eqref{eq:equation_u} - \eqref{eq:equation_RttRrr} with
the boundary conditions \eqref{eq:BC_origin} - \eqref{eq:BC_infinity_metric} using a numerical relaxation method and discuss the physical properties.
In this section, we fix the parameters as $e=0.1$, $\mu=1.4$,
$\lambda=1.0$, and $G\eta^2=\left(\eta/M_P\right)^2=10^{-4}$.
In the next section, we obtain solutions in the case of different values of
$\lambda$, and $G\eta^2$.

\subsection{Field configurations}
\begin{figure}[H]
\begin{tabular}{ccc}
\begin{minipage}{0.33\hsize}
\centering
\text{(d) $\Omega=0.828$}\par\medskip
\includegraphics[width=\figwidth]{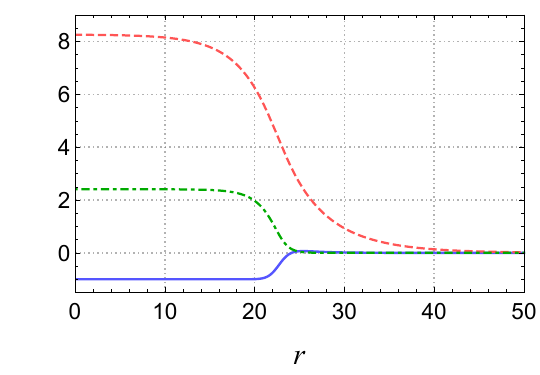}\\
\includegraphics[width=\figwidth]{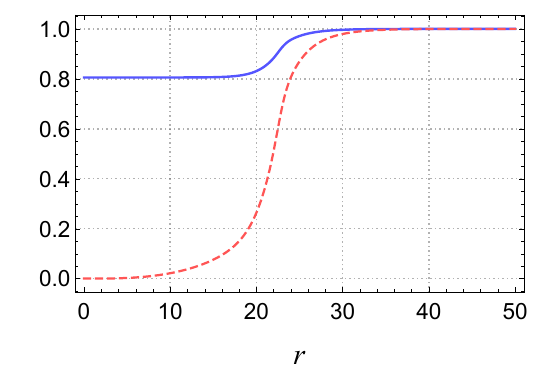}
\end{minipage}
\begin{minipage}{0.33\hsize}
\centering
\text{(e) $\Omega=0.750$}\par\medskip
\includegraphics[width=\figwidth]{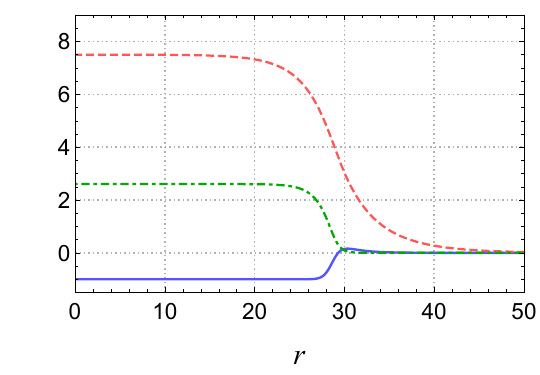}\\
\includegraphics[width=\figwidth]{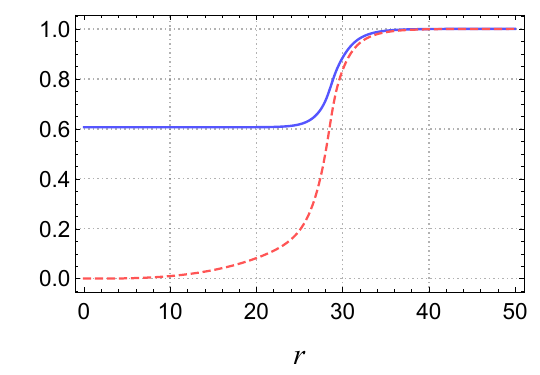}
\end{minipage}
\begin{minipage}{0.33\hsize}
\centering
\text{(f) $\Omega=0.665$}\par\medskip
\includegraphics[width=\figwidth]{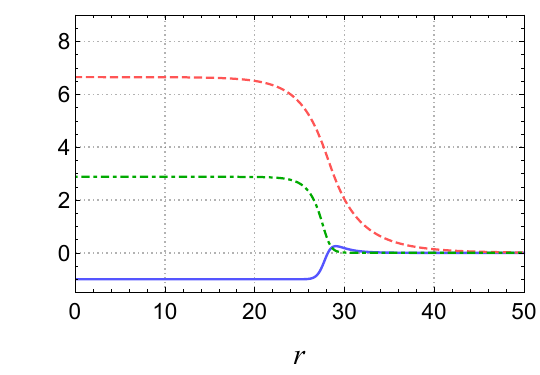}\\
\includegraphics[width=\figwidth]{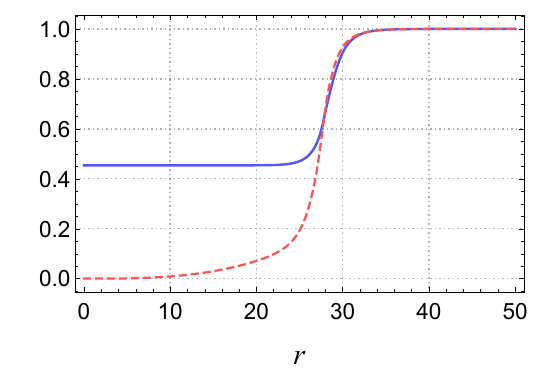}
\end{minipage}
\\
\end{tabular}
\caption{
Numerical solutions for $\lambda=1.0$ and $\eta/M_P=10^{-2}$ are depicted for
(a) $\Omega=0.750$ (upper-left column), (b) $\Omega=0.677$ (upper-middle column), (c) $\Omega=0.750$ (upper-right column), (d) $\Omega=0.828$ (lower-left column), (e) $\Omega=0.750$ (lower-middle column), and (f) $\Omega=0.665$ (lower-left column).
The complex scalar fields $f,u$ and the gauge field $\alpha$ are plotted in each upper panel.
The lapse function $\sigma$ and the mass function $m$ are plotted in each lower panel.
\label{fig:configuration_lambda1_G10e4}
}
\end{figure}

Numerical solutions for various $\Omega$ are plotted in Fig.\ref{fig:configuration_lambda1_G10e4}.
In all cases (a) - (f), the fields $u,f$, and $\alpha$ are excited in a finite spatial region,
decay quickly
and approach to vacuum state \eqref{eq:BC_infinity_matter} in far region.
These numerical solutions (a) - (f) indicate nontopological solitons
with self-gravity, we call them nontopological soliton star (NTS star).
In particular, in the cases (d) - (f), the variables take constant values in the central
region of $r$, and decay within a layer of finite thickness.
We call the layer the surface layer.
These solutions are self-gravitating potential balls discussed in ref.	\cite{Ishihara:2021iag}.
In the interior region surrounded by the surface layer,
the constants that $f$, and $\alpha$ take are
\begin{align}
	f=0~~,~~\alpha=\frac{\Omega}{e}.
\label{eq:f0alpha0u0}
\end{align}

\begin{figure}[H]
\begin{tabular}{ccc}
\begin{minipage}{0.33\hsize}
\centering
\text{(a) $\Omega=0.750$}\par\medskip
\includegraphics[width=\figwidth]{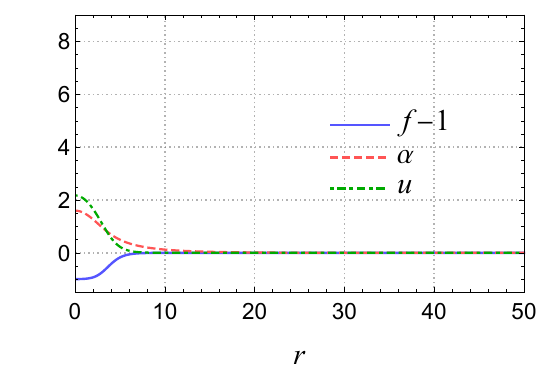}\\
\includegraphics[width=\figwidth]{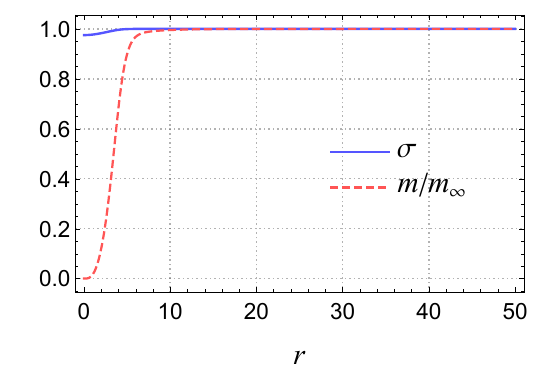}
\end{minipage}
\begin{minipage}{0.33\hsize}
\centering
\text{(b) $\Omega=0.677$}\par\medskip
\includegraphics[width=\figwidth]{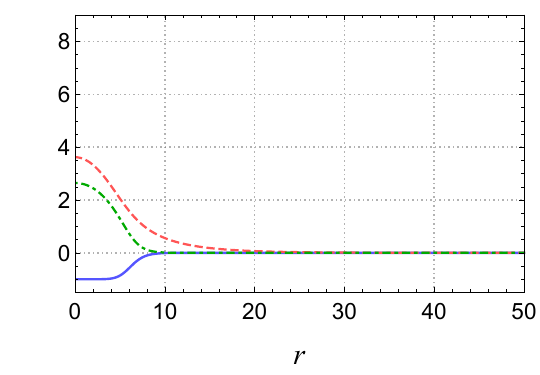}\\
\includegraphics[width=\figwidth]{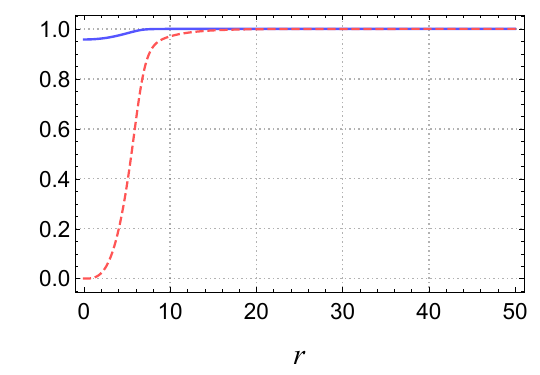}
\end{minipage}
\begin{minipage}{0.33\hsize}
\centering
\text{(c) $\Omega=0.750$}\par\medskip
\includegraphics[width=\figwidth]{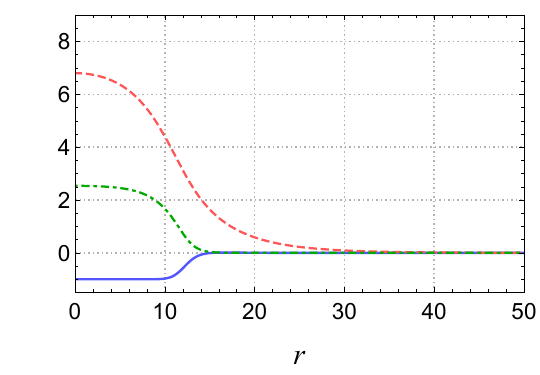}\\
\includegraphics[width=\figwidth]{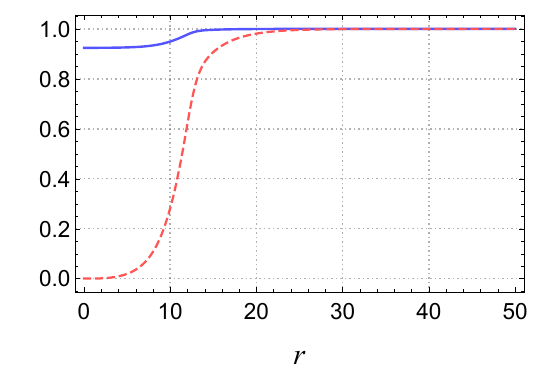}
\end{minipage}
\\
\\
\end{tabular}
\\
\\
\end{figure}

The lapse function $\sigma$ takes nearly 1 for (a) - (c), while $\sigma$ is clearly less than 1 for
(d) - (f).
It means that the gravity is inefficient in the cases (a) - (c), while efficient in
(d) - (f).
In the exterior of the surface layer, the metric functions take $\sigma=1$ and $m=m_{\infty}=const.$
This means that the geometry of the exterior vacuum region is given by the Schwarzschild metric
\begin{align}
  ds^2=&-\left(1-\frac{2m_{\infty}}{r}\right)dt^2+\left(1-\frac{2m_{\infty}}{r}\right)^{-1}dr^2
		+r^2(d\theta^2+\sin^2\theta d\varphi^2).
 \label{eq:effective_metric_outside}
\end{align}

\subsection{Energy density and Pressure}
Energy density $\epsilon$, radial pressure $p_{\perp}$, and tangential pressure $p_{\parallel}$ for the cases (a) - (f) are plotted in Fig.\ref{fig:G10e4_EnergyPressure}.
The definitions of $\epsilon,p_{\perp}$, and $p_{\parallel}$ are presented in Appendix \ref{appendixA}.

\begin{figure}[H]
\begin{tabular}{ccc}
\begin{minipage}{0.33\hsize}
\centering
\text{(a) $\Omega=0.750$}\par\medskip
\includegraphics[width=\figwidth]{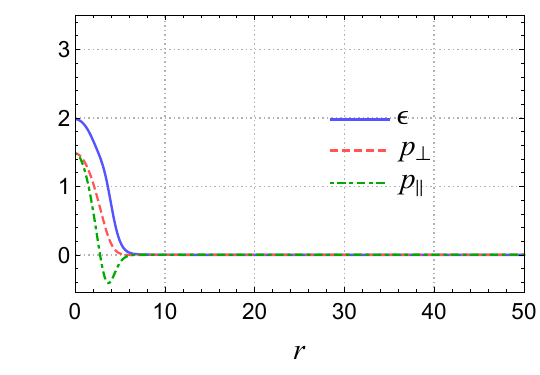}
\end{minipage}
\begin{minipage}{0.33\hsize}
\centering
\text{(b) $\Omega=0.677$}\par\medskip
\includegraphics[width=\figwidth]{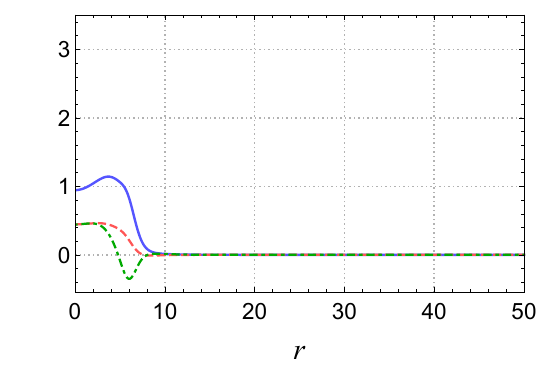}
\end{minipage}
\begin{minipage}{0.33\hsize}
\centering
\text{(c) $\Omega=0.750$}\par\medskip
\includegraphics[width=\figwidth]{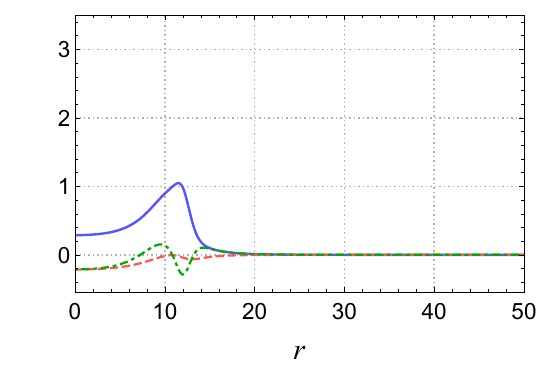}
\end{minipage}
\\
\vspace{5mm}
\end{tabular}
\\

\begin{tabular}{ccc}
\begin{minipage}{0.33\hsize}
\centering
\text{(d) $\Omega=0.828$}\par\medskip
\includegraphics[width=\figwidth]{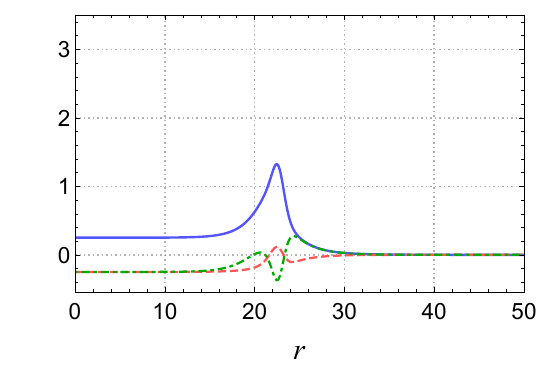}
\end{minipage}
\begin{minipage}{0.33\hsize}
\centering
\text{(e) $\Omega=0.750$}\par\medskip
\includegraphics[width=\figwidth]{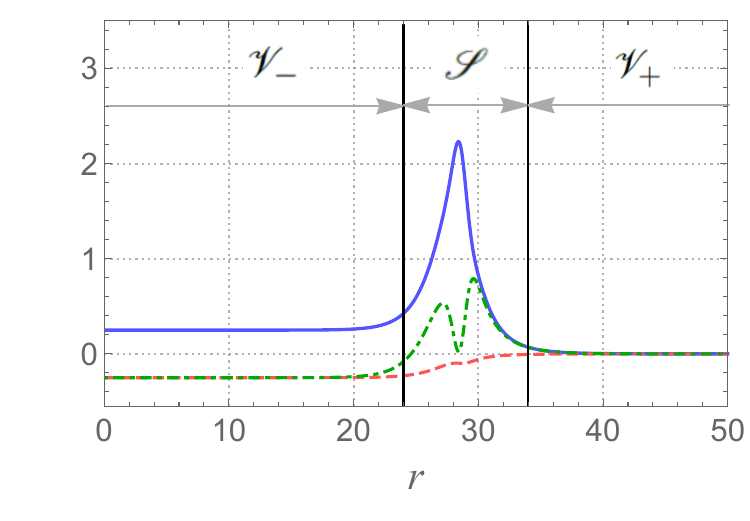}
\end{minipage}
\begin{minipage}{0.33\hsize}
\centering
\text{(f) $\Omega=0.665$}\par\medskip
\includegraphics[width=\figwidth]{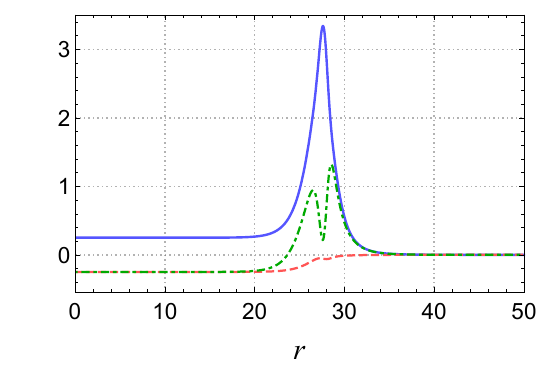}
\end{minipage}
\\
\end{tabular}
\caption{
Energy density, $\epsilon$, radial pressure, $p_{\perp}$, and tangential pressure, $p_{\parallel}$ for the NTS stars are plotted as functions of $r$. The values of $\Omega$ are same as (a) - (f) in Fig.\ref{fig:configuration_lambda1_G10e4}.
}
\label{fig:G10e4_EnergyPressure}
\end{figure}

For the solutions (d) - (f), we define three regions (see the center of lower panel in Fig.\ref{fig:G10e4_EnergyPressure})
\begin{align*}
  \mathscr{V}_{-}~&:~ \text{interior region surrouded by the surface layer,}\\
  \mathscr{S}~&:~ \text{inside the thickness of the surface layer,}\\
  \mathscr{V}_{+}~&:~ \text{exterior region of the surface layer.}
\end{align*}
The energy density $\epsilon$ has a sharp peak in $\mathscr{S}$.
We define the surface layer radius, say $r_{\rm sl}$,
such that $\epsilon$ is maximum at the radius.

In the region $\mathscr{V}_-$, it is common for (d) - (f) in Fig.\ref{fig:G10e4_EnergyPressure}
that $\epsilon$ takes a constant, $\epsilon_0=0.25$, and
$p_{\parallel}=p_{\perp}=-\epsilon_0$.
In the region $\mathscr{S}$, $p_{\parallel}$ has double peaks, and $p_{\perp}$ is
negligibly small compare to $p_{\parallel}$.
In the region $\mathscr{V}_+$, $\epsilon, p_{\parallel}$, and $p_{\perp}$ decay exponentially
to zero as $r$ increases.
In $\mathscr{V}_-$, substituting the field values \eqref{eq:f0alpha0u0}
into \eqref{eq:dimensionless_Ttt} - \eqref{eq:dimensionless_Tthetatheta},
we find that the energy-momentum tensor reduces to the potential term of $\phi$,
i.e.,
\begin{align}
	T^t_{~t}=T^r_{~r}=T^\theta_{~\theta}=T^\varphi_{~\varphi}=-\frac{\lambda}{4}.
  \label{eq:EMtensor_gravastar}
\end{align}
Therefore, the equation of state of the sources of gravity in $\mathscr{V}_-$ are given by
\begin{align}
  \epsilon=-p_{\perp}=-p_{\parallel}=\epsilon_0.
  \label{eq:EOS_Interior}
\end{align}
From Fig.\ref{fig:G10e4_EnergyPressure}, we see the constant value $\epsilon_0$ is coincided
with $\lambda/4=0.25$.
In the region $\S$, $\epsilon, p_{\perp}$, and $p_{\parallel}$ are supplied mainly
by the energy-momentum tensor of electric field and the kinetic energy of the scalar fields
given by \eqref{eq:dimensionless_Ttt} - \eqref{eq:dimensionless_Tthetatheta}.

In summary, for the solutions (d) - (f), the interior region $\mathscr{V}_-$
with the non-vanishing vacuum energy \eqref{eq:EOS_Interior} and the exterior region
$\mathscr{V}_+$ of the true vacuum are attached by the surface layer $\S$.
This properties are the same as so-called gravastars \cite{Mazur:2001fv,Mazur:2004fk}.
From these results, we call the NTS stars (d) - (f) \lq solitonic gravastars'.

\begin{figure}[H]
\centering
\includegraphics[width=7cm]{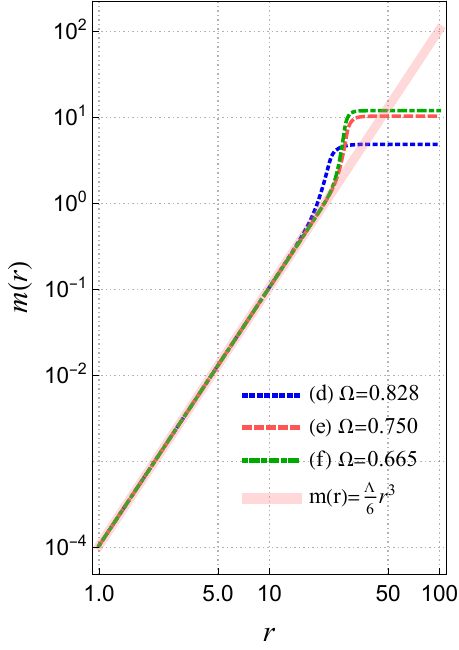}
\caption{
The mass functions $m(r)$ by logarithmic scale for the cases $\Omega=0.828,~0.800$, and $0.665$.
The thick solid line that denotes $m(r)=(\Lambda/6)r^3$ is shown as a reference, where the value of $\Lambda \sim 6.3\times 10^{-4}$ is given by
\eqref{eq:value_Lambda}.
}
\label{fig:G10e4_loglogmass}
\end{figure}

\subsection{Behavior of mass function $m$}
In Fig.\ref{fig:G10e4_loglogmass}, we show behaviors of the mass functions $m(r)$
by logarithmic scale for the solitonic gravastars (d) - (f).
We can see that $m(r)=6.3\times 10^{-4} r^3$
in $\mathscr{V}_-$ in all cases,
increases significantly in $\mathscr{S}$.
In $\mathscr{V}_+$, $m(r)$ take constant values $m_{\infty}$ that depends on $\Omega$.

In $\mathscr{V}_-$, substituting the relation \eqref{eq:EMtensor_gravastar} into \eqref{eq:equation_Rtt}, we reduce time-time component of the Einstein equation to
\begin{align}
  &\frac{2m'}{r^2}-\Lambda=0,
\cr
	&\Lambda:=8\pi G\eta^2\frac{\lambda}{4}.
\end{align}
Therefore we obtain
\begin{align}
  m(r)=\frac{\Lambda}{6} r^3,
  \label{eq:m_inside}
\end{align}
where the value of $\Lambda$ is given by
\begin{align}
	\Lambda
	\sim 6.3\times 10^{-4}
\label{eq:value_Lambda}
\end{align}
for the fixed values of parameters $\lambda=0.1$ and $G\eta^2=10^{-4}$.
From Fig.\ref{fig:G10e4_loglogmass}, we can see the value of $\Lambda$ is coincided
with numerical solutions.
Thus, the metric in $\mathscr{V}_-$ has the form of de Sitter spacetime as
\begin{align}
  ds^2=&-\left(1-\frac{\Lambda}{3} r^2\right){d\tilde{t}}^2
	+\left(1-\frac{\Lambda}{3} r^2\right)^{-1}dr^2
	+r^2 (d\theta^2+\sin^2\theta d\varphi^2),
 \label{eq:effective_metric1}
\end{align}
where $\tilde{t}:=\sigma_0 t$.

\subsection{Characteristic radii of the solitonic gravastars}

For the solitonic gravastar solutions (d) - (f), in addition to the surface layer radius, $r_{\rm sl}$, there are two characteristic radii:
the Schwarzschild radius, $r_{\rm Sch}$,
and the de Sitter horizon radius, $r_{\rm dS}$, defined by
\begin{align}
  r_{\rm Sch} := 2m_\infty, \quad \mbox{and}\quad
  r_{\rm dS} := \sqrt{\frac{3}{\Lambda}}.
\end{align}
In Table \ref{table:various_radius}, we summarized the values of radii for the cases (d) - (f).
Therefore we have the relation $r_{\rm Sch} <r_{\rm sl} <r_{\rm dS}$, namely,
there is neither a Schwarzschild horizon nor a de Sitter horizon.

\begin{table}[htb]
  \begin{tabular}{c|c|c|c}
    \hline
      & $r_{\rm Sch}$ & $r_{\rm sl}$ & $r_{\rm dS}$ \\ \hline \hline
    $(d)~\Omega=0.828$ & $~~10~~$ & $~~22~~$ & $~~70~~$ \\ \hline
    $(e)~\Omega=0.750$ & $~~16~~$ & $~~28~~$ & $~~70~~$ \\ \hline
    $(f)~\Omega=0.665$ & $~~24~~$ & $~~28~~$ & $~~70~~$ \\
    \hline
  \end{tabular}
\caption{
The values of the Schwarzschild radius, $r_{\rm Sch}$, the surface layer radius, $r_{\rm sl}$, and the de Sitter horizon radius, $r_{\rm dS}$, in the unit of $\eta=1$ for the solitonic gravastar solutions (d) - (f).
}
  \label{table:various_radius}
\end{table}

\subsection{Mass of the NTS stars}
\begin{figure}[H]
\centering
\includegraphics[width=10cm]{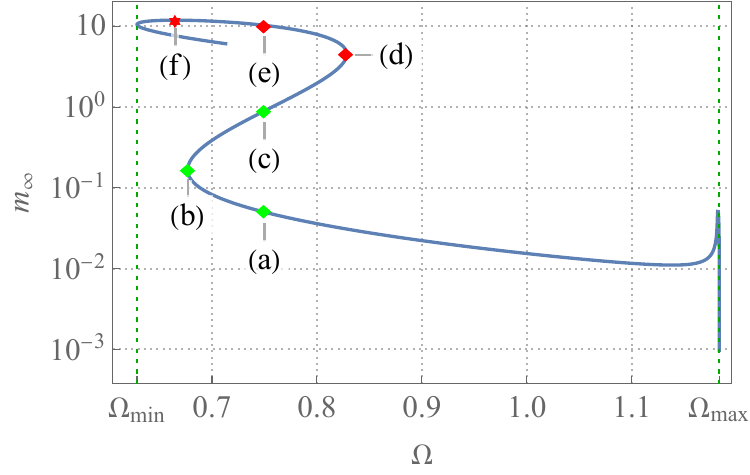}
\caption{
Mass at infinity $m_{\infty}$ is plotted as a function of $\Omega$.
Diamonds in the figure correspond to (a), (c), (e) $\Omega=0.750$, (b) $\Omega=0.677$, and (d)  $\Omega=0.828$.
The asterisk, which corresponds to (f) $\Omega_{\ast}=0.665$, indicates maximum value of $m_{\infty}$.
The red marks are solitonic gravastars.
}
\label{fig:G10e4_Omega_Totalmass}
\end{figure}

We see, in Fig.\ref{fig:G10e4_Omega_Totalmass}, that
$m_\infty(\Omega)$ is a multi-valued function of $\Omega$.
The mass $m_\infty$ takes maximum value, $m_{\ast}$,
for the solitonic gravastar solution (f) with $\Omega=\Omega_*=0.665$.
We call this solution with $(m_{\ast}, \Omega_*)$ the \lq maximum gravastar'.
We find the solutions exist on a curve in the range
\begin{align}
	\Omega_{min} \leq \Omega < \Omega_{max}.
\label{eq:Omega_range}
\end{align}
In the limit $r\to \infty$, from the boundary conditions
\eqref{eq:BC_origin} - \eqref{eq:BC_infinity_metric},
we require $\sigma\to 1,~F\to 1$, and $\alpha\to0,~f\to 1$,
then, \eqref{eq:equation_u} reduces to
\begin{align}
  u''-(\mu - \Omega^2 ) u = 0.
\label{large_lim}
\end{align}
Therefore, in order that $u$ decays exponentially, upper bound $\Omega_{max}$ is given
by $\Omega_{max}:=\sqrt{\mu}$.

\subsection{Charge density}
\begin{figure}[H]
\begin{tabular}{ccc}
\begin{minipage}{0.33\hsize}
\centering
\text{(d) $\Omega=0.828$}\par\medskip
\includegraphics[width=\figwidth]{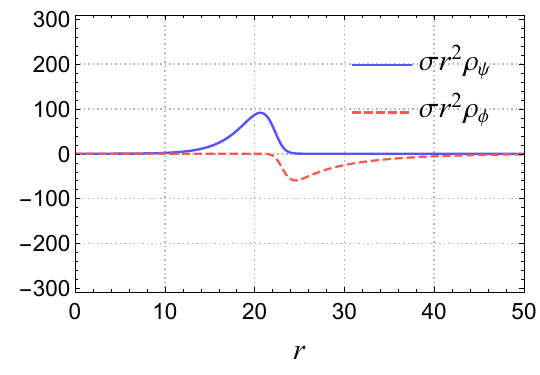}
\end{minipage}
\begin{minipage}{0.33\hsize}
\centering
\text{(e) $\Omega=0.750$}\par\medskip
\includegraphics[width=\figwidth]{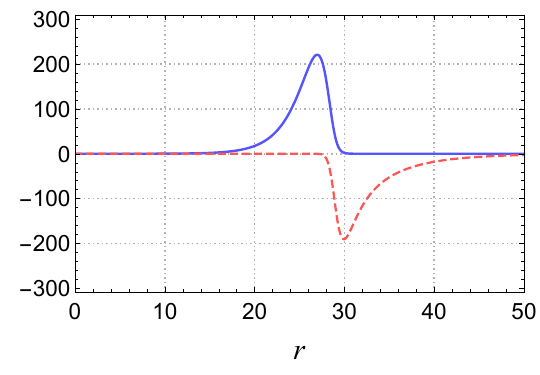}
\end{minipage}
\begin{minipage}{0.33\hsize}
\centering
\text{(f) $\Omega_{\ast}=0.665$}\par\medskip
\includegraphics[width=\figwidth]{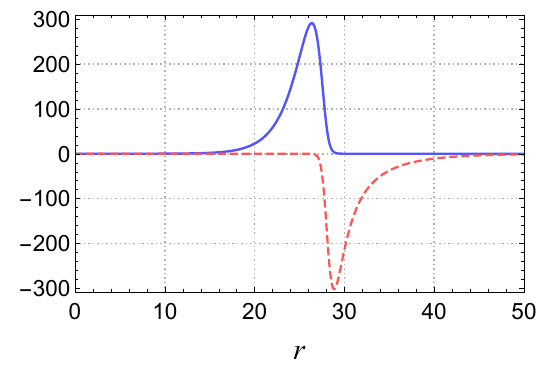}
\end{minipage}
\\
\end{tabular}
\caption{
The charge densities $\rho_{\psi}$ and $\rho_{\phi}$ of the complex scalar fields $\psi$ and $\phi$, respectively for the solitonic gravastars (d) - (f).
}
\label{fig:G10e4_chargedensity}
\end{figure}

The charge densities $\rho_{\psi}$ and $\rho_{\phi}$ of the complex scalar fields $\psi$
and $\phi$ appear.
In Fig.\ref{fig:G10e4_chargedensity}, we see that $\rho_{\psi}=\rho_{\phi}=0$
in the both regions $\mathscr{V}_-$ and $\mathscr{V}_+$, and
$\rho_{\psi}>0~,~\rho_{\phi}=0$ in the inner edge of $\mathscr{S}$, while
$\rho_{\psi}=0~,~\rho_{\phi}<0$ in the outer edge of $\mathscr{S}$.
Namely, an electric double layer emerges in $\mathscr{S}$.

Owing to the boundary conditions \eqref{eq:BC_infinity_matter},
the gauge field decays exponentially as $r$ increases to spatial infinity.
Then, for the solitonic gravastars in our model, the relation
\begin{align}
  4\pi\int_0^{\infty}\sigma r^2\rho_{\psi}dr+4\pi\int_0^{\infty}\sigma r^2\rho_{\phi}dr=0
\end{align}
should be satisfied.
This means that $\rho_{\psi}$ and $\rho_{\phi}$ are totally screened each other.

\subsection{Radius and Compactness}
In this subsection, we discuss the compactness of the NTS stars.
First, let us define the radius of NTS stars, $r_{\rm s}$, by
\begin{align}
	m(r_{\rm s}):=0.99~ m_{\infty},
\label{eq:numerical_R}
\end{align}
namely, 99\% of total mass for the NTS stars is included within the radius $r_{\rm s}$.
The radius $r_{\rm s}$ is slightly larger than $r_{\rm sl}$.

\begin{figure}[H]
\centering
\includegraphics[width=10cm]{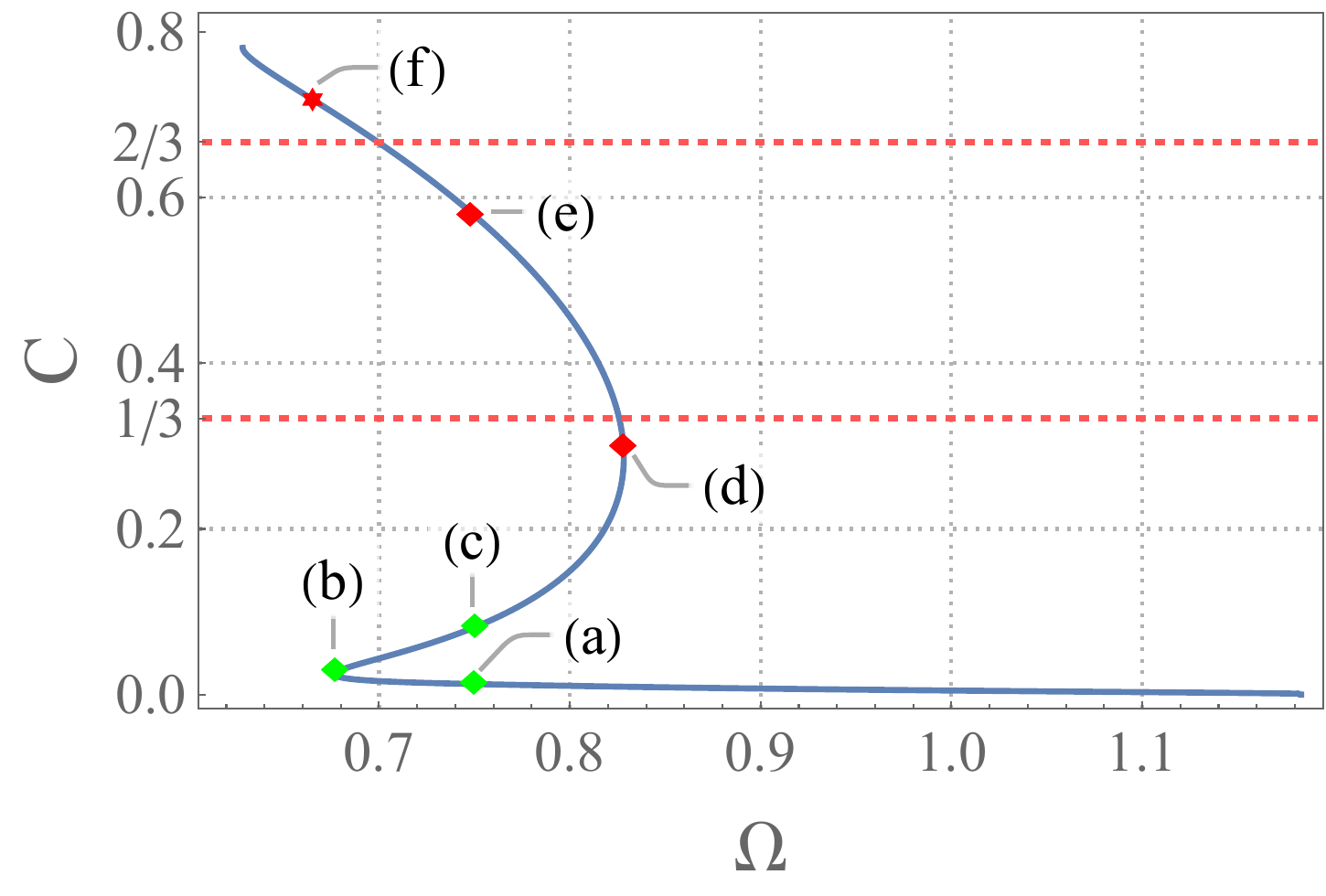}
\caption{
Compactness, $C$, is plotted as a function of $\Omega$.
The dashed lines indicate the lower limits of $C$ for existence of ISCO $(C=1/3)$ and photon sphere $(C=2/3)$, respectively.
The red marks (d), (e), (f) are solitonic gravastars, and the asterisk corresponds to the maximum gravastar.
}
\label{fig:G10e4_Omega_Compactness}
\end{figure}

Using $r_{\rm s}$ and $m_\infty$, we define compactness $C$ for the NTS stars by
\begin{align}
  C:=\frac{2m_{\infty}}{r_{\rm s}}.
\end{align}
In the Schwarzschild metrics, which describe the geometry in $\mathscr{V}_+$,
the innermost stable circular orbit (ISCO) appears if $C\ge 1/3$, and
both the ISCO and the photon sphere exist if $C\ge 2/3$.
The compactness of NTS stars as a function of $\Omega$ is plotted
in Fig.\ref{fig:G10e4_Omega_Compactness}.
The NTS stars (a), (b), (c), and the solitonic gravastar (d), have
neither an ISCO nor a photon sphere.
The solitonic gravastar (e) has an ISCO but no photon sphere.
The maximum gravastar (f) has both an ISCO and a photon sphere.
Solitonic gravastars with various compactness can exist.

\section{Maximum gravastars for various sets of $(\lambda,\eta/M_P)$}

As shown in Fig.\ref{fig:G10e4_Omega_Totalmass}, for a fixed set of parameters
$(e,\mu,\lambda,\eta/M_P)$, the maximum gravastar is obtained
by fine tuning of $\Omega$.
In this section, we obtain the maximum gravastars for other sets of parameters $(\lambda,\eta/M_P)$,
and study how the dimension-full gravitational mass defined by
\begin{align}
  M_G := \frac{m_\infty}{G\eta^2}\eta
 \label{eq:gravitational_mass}
\end{align}
depends on the parameters.

\subsection{Variation of $\eta/M_P$}
\begin{figure}[H]
\centering
\includegraphics[width=10cm]{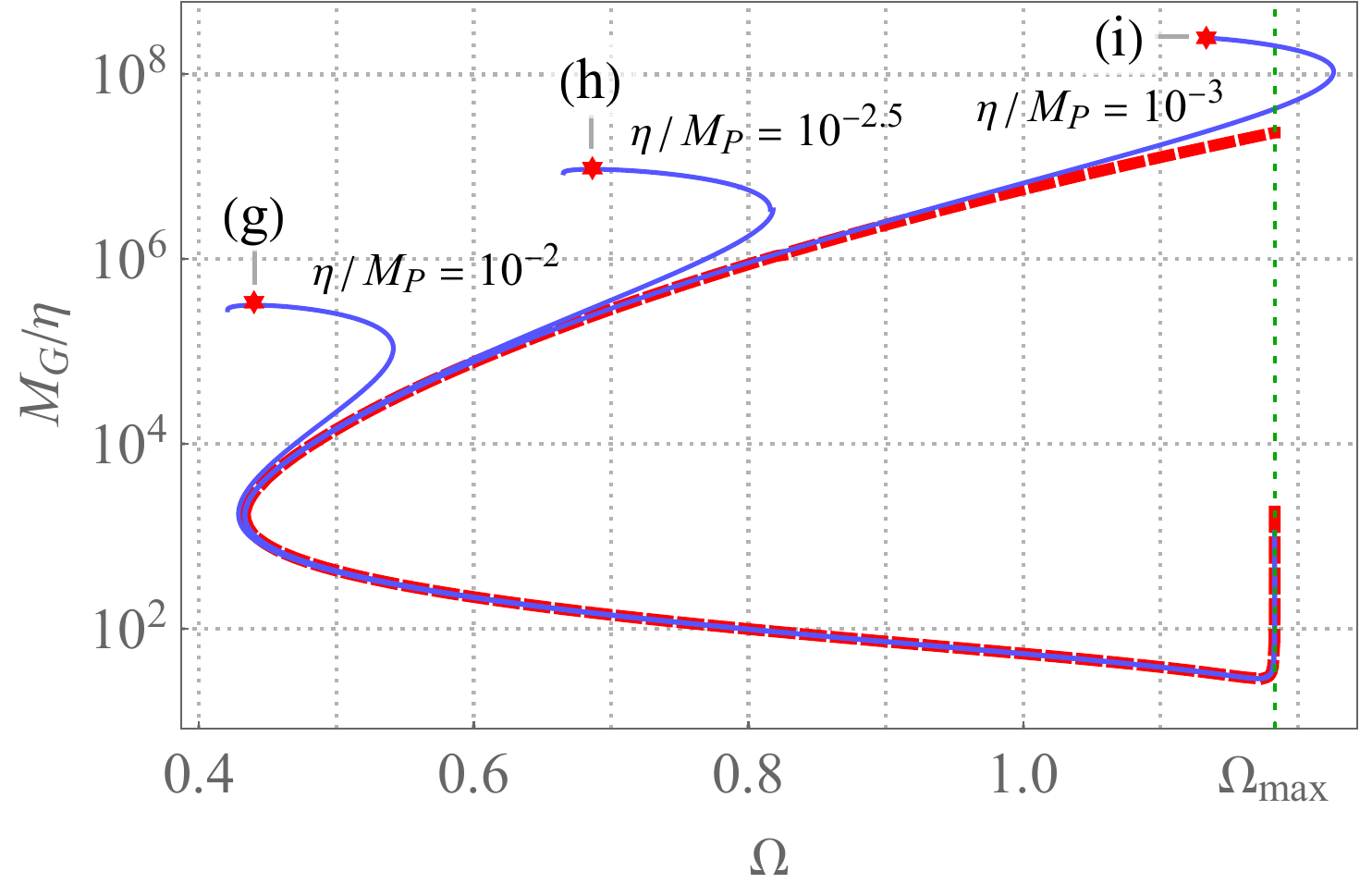}
\caption{
Gravitational mass $M_G$ is plotted as functions of $\Omega$ for $\lambda=0.1$ and $\eta/M_P=10^{-2},~10^{-2.5}$, and $10^{-3}$, respectively.
The asterisks indicate maximum gravastars for each value of $\eta/M_P$.
The dashed line represents total mass in the limit $\eta/M_P \to 0$, non-gravitating NTSs.
}
\label{fig:lambda0p1_variouseta_Omega_Totalmass}
\end{figure}

\begin{figure}[H]
\begin{tabular}{ccc}
\begin{minipage}{0.33\hsize}
\centering
\text{(g) $\eta/M_P=10^{-2}~,~\Omega_{\ast}=0.443$}\par\medskip
\includegraphics[width=\figwidth]{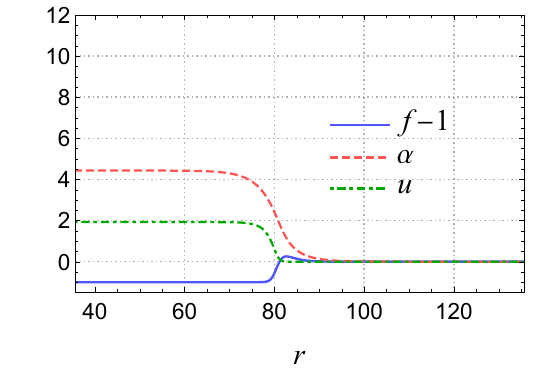}\\
\includegraphics[width=\figwidth]{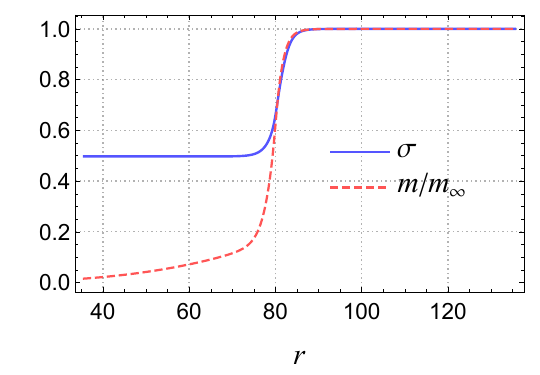}
\end{minipage}
\begin{minipage}{0.33\hsize}
\centering
\text{(h) $\eta/M_P=10^{-2.5}~,~\Omega_{\ast}=0.686$}\par\medskip
\includegraphics[width=\figwidth]{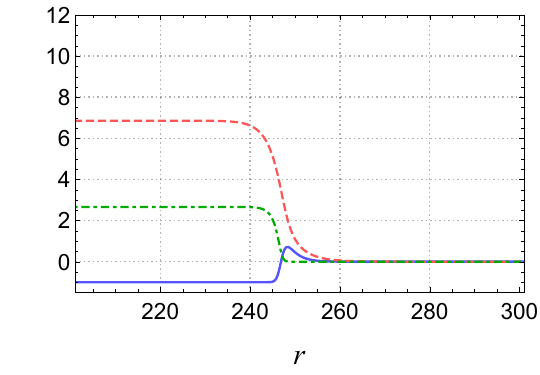}\\
\includegraphics[width=\figwidth]{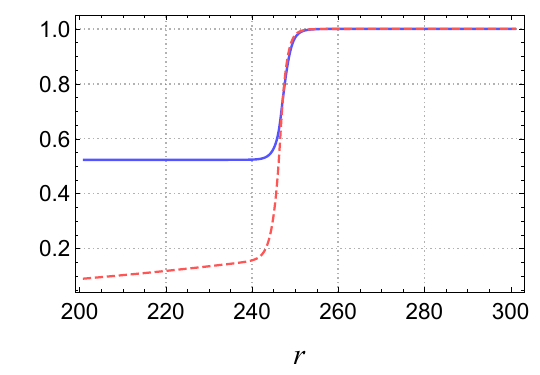}
\end{minipage}
\begin{minipage}{0.33\hsize}
\centering
\text{(i) $\eta/M_P=10^{-3}~,~\Omega_{\ast}=1.134$}\par\medskip
\includegraphics[width=\figwidth]{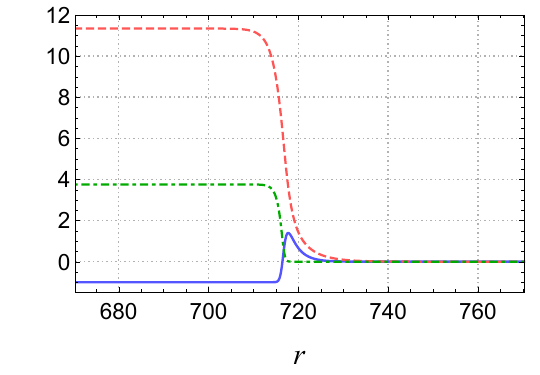}\\
\includegraphics[width=\figwidth]{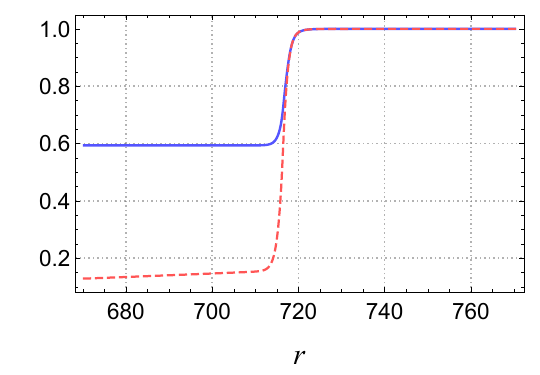}
\end{minipage}
\\
\end{tabular}
\caption{
The maximum gravastar solutions for $\lambda=0.1$ are depicted for (g) $\eta/M_P=10^{-2},~\Omega_{\ast}=0.443$ (left column), (h) $\eta/M_P=10^{-2.5},~\Omega_{\ast}=0.686$ (middle column), (i) $\eta/M_P=10^{-3},~\Omega_{\ast}=1.134$ (right column).
The complex scalar fields $f,u$ and the gauge field $\alpha$ are plotted in upper panels.
The lapse function $\sigma$ and the mass function $m$ are plotted as functions of $r$ in lower panels.
\label{fig:Mmax_configuration_lambda_0p1_various__G}
}
\end{figure}

We display $M_G$ as a function of $\Omega$ for $\lambda=0.1$ and
$\eta/M_P=10^{-2},~10^{-2.5}$, and $10^{-3}$
in Fig.\ref{fig:lambda0p1_variouseta_Omega_Totalmass}.
In the range $M_G/\eta\lesssim (\eta/M_P)^{-2}$ of each case (g), (h), (i),
the curve $M_G(\Omega)$ follows the limiting curve of $\eta/M_P \to 0$,
i.e., curve of non-gravitating NTS.
(See Fig.\ref{fig:lambda0p1_variouseta_Omega_Totalmass}).
The each curve $M_G(\Omega)$ moves away from the limiting curve
in the region $M_G/\eta \gtrsim (\eta/M_P)^{-2}$.
We find that the values of $\Omega_{\ast}$ and $M_*$ increase
as $\eta/M_P$ decreases.
In the case of $\eta/M_P=10^{-3}$, solutions with $\Omega>\Omega_{max}$ appear.
In such solutions, described by \eqref{large_lim} in the large $r$ region,
$u$ does not decay exponentially but oscillates
as $r$ increases toward infinity.

The configurations of variables for the maximum gravastars in the cases,
$\lambda=0.1$ and $\eta/M_P=10^{-2},~10^{-2.5}$, and $10^{-3}$ are plotted
in Fig.\ref{fig:Mmax_configuration_lambda_0p1_various__G}.
In all cases (g), (h), and (i), values of $f$ and $\alpha$ satisfy \eqref{eq:f0alpha0u0}
in the region $\mathscr{V}_-$, respectively.
We note that overshooting of $f$, which appears near $r_{\rm sl}$,
becomes large as $\eta/M_P$ decreases.

The radius of the surface layer $r_{\rm sl}$ increases as $\eta/M_P$ decrease
while the thickness of the surface layer can be estimate by $\Delta r\sim 10\eta^{-1}$
for all cases (g), (h), and (i).
Then, the ratio $\Delta r/r_{\rm sl}$ decreases as $\eta/M_P$ decreases.
Thus, the solitonic gravastars with small $\eta/M_P$ could be described
by
junction of de Sitter and Schwarzschild spacetimes with a thin-shell.

\subsection{Variation of $\lambda$}

\begin{figure}[H]
\centering
\includegraphics[width=10cm]{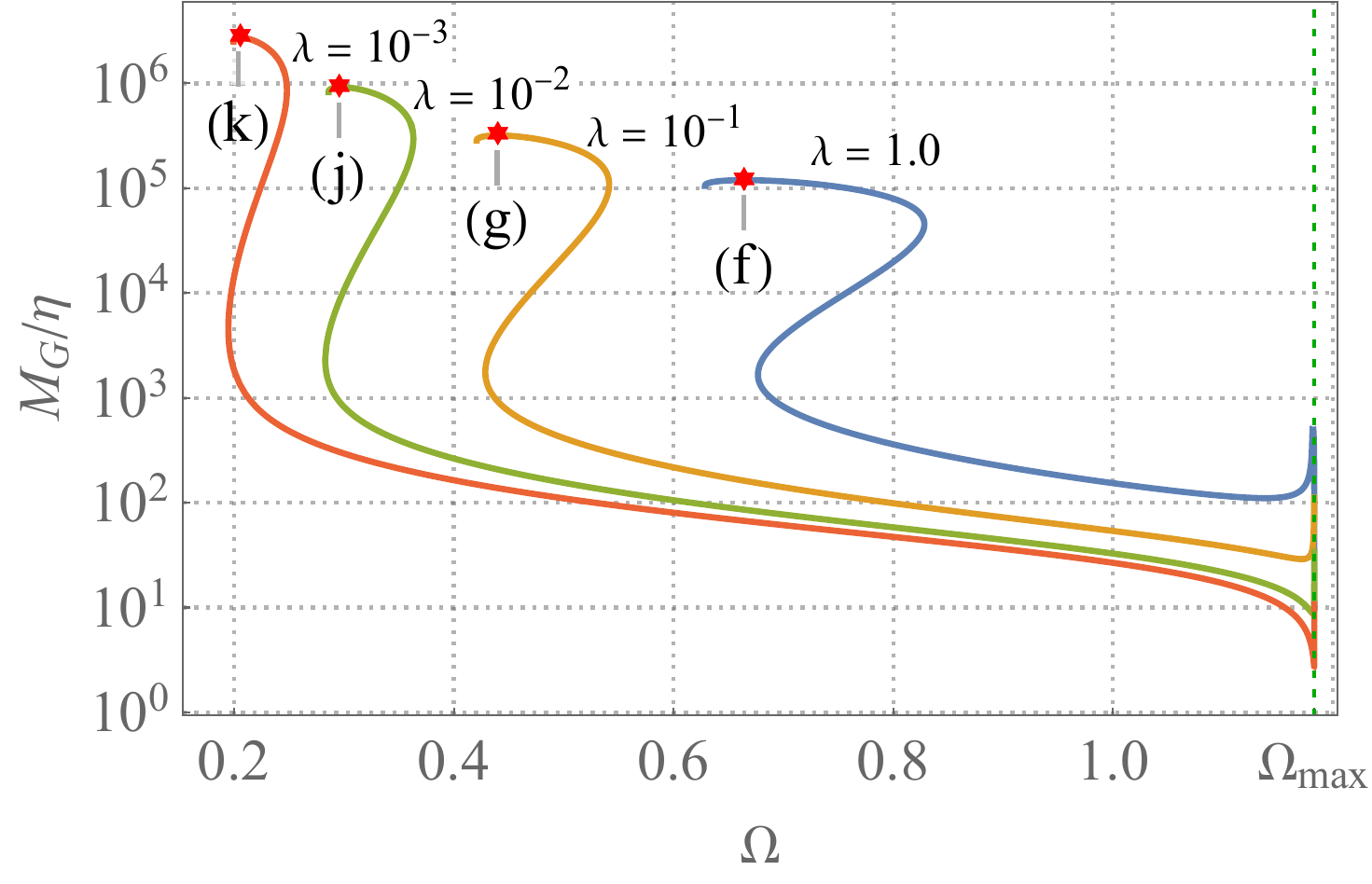}
\caption{
Gravitational mass $M_G$ is plotted for the cases of $\eta/M_P=10^{-2}$ and
$\lambda=1.0,~10^{-1},~10^{-2},~10^{-3}$.
The asterisks indicate the maximum gravastar for each value of $\lambda$.
}
\label{fig:G10e4_variouslambda_Omega_Totalmass}
\end{figure}
We display $M_G$ as a function of $\Omega$ for fixed $\eta/M_P=10^{-2}$ and
$\lambda=1.0,~10^{-1},~10^{-2}$, and $10^{-3}$
in Fig.\ref{fig:G10e4_variouslambda_Omega_Totalmass}.
The values of $M_*$ increase and $\Omega_{\ast}$ decreases as $\lambda$ decreases.

\begin{figure}[H]
\begin{tabular}{cccc}
\begin{minipage}{0.25\hsize}
\centering
\text{(f) $\lambda=1.0$~,~}\par
\hspace{0.8cm} \text{$\Omega_{\ast}=0.665$}\par\medskip
\includegraphics[width=4.5cm]{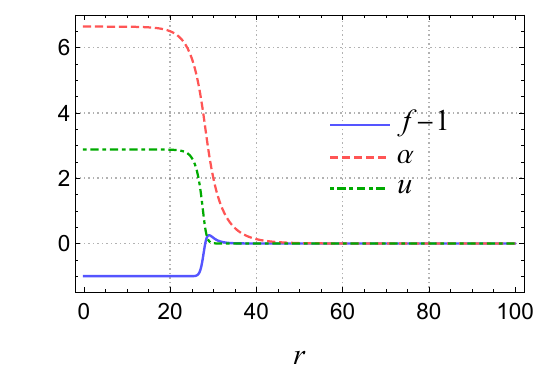}\\
\includegraphics[width=4.5cm]{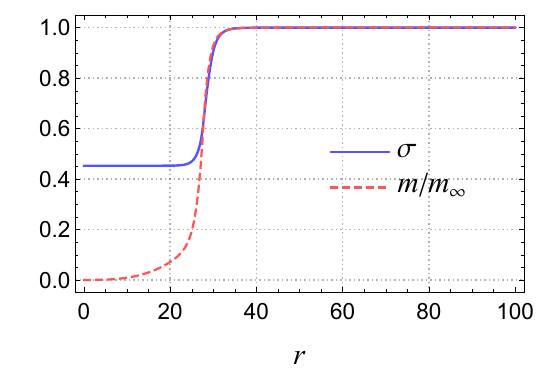}
\end{minipage}
\begin{minipage}{0.25\hsize}
\centering
\text{(g) $\lambda=10^{-1}$~,~}\par
\hspace{0.7cm} \text{$\Omega_{\ast}=0.443$}\par\medskip
\includegraphics[width=4.5cm]{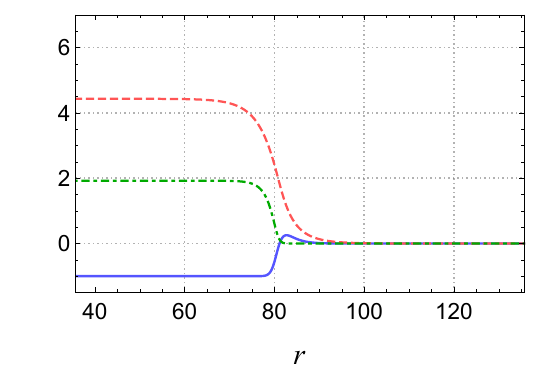}\\
\includegraphics[width=4.5cm]{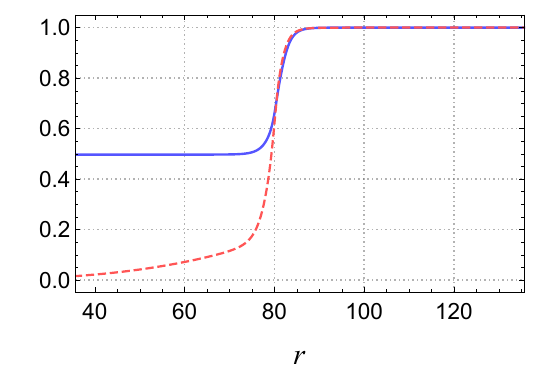}
\end{minipage}
\begin{minipage}{0.25\hsize}
\centering
\text{(j) $\lambda=10^{-2}$~,~}\par
\hspace{0.6cm} \text{$\Omega_{\ast}=0.297$}\par\medskip
\includegraphics[width=4.5cm]{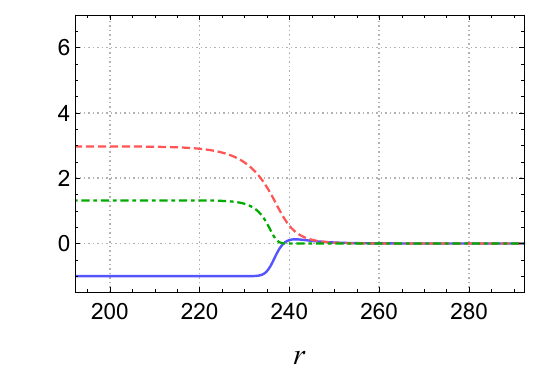}\\
\includegraphics[width=4.5cm]{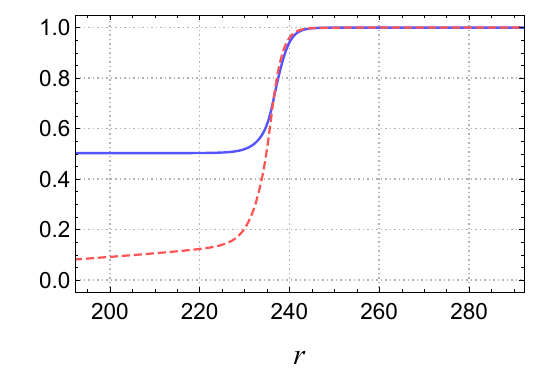}
\end{minipage}
\begin{minipage}{0.25\hsize}
\centering
\text{(k) $\lambda=10^{-3}$~,~}\par
\hspace{0.5cm} \text{$\Omega_{\ast}=0.206$}\par\medskip
\includegraphics[width=4.5cm]{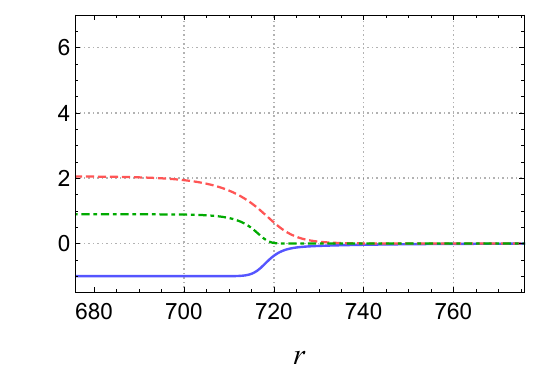}\\
\includegraphics[width=4.5cm]{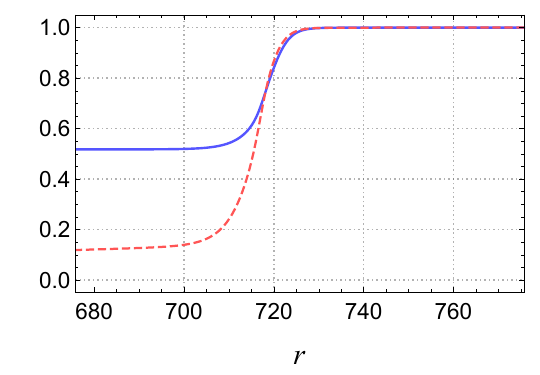}
\end{minipage}
\\
\end{tabular}
\caption{
The maximum gravastar solutions for $\eta/M_P=10^{-2}$ are depicted for (f) $\lambda=1.0,~\Omega_{\ast}=0.665$, (g) $\lambda=10^{-1},~\Omega_{\ast}=0.443$, (j) $\lambda=10^{-2},~\Omega_{\ast}=0.297$, (k) $\lambda=10^{-3},~\Omega_{\ast}=0.206$.
The complex scalar fields $f,u$ and the gauge field $\alpha$ are plotted in upper panels.
The lapse function $\sigma$ and the mass function $m$ are plotted as functions of $r$ in lower panels.
\label{fig:Mmax_configuration_G_10e4_various_lambda}
}
\end{figure}
The configurations of variables for the maximum gravastars in these cases,
are plotted in Fig.\ref{fig:Mmax_configuration_G_10e4_various_lambda}.
In all cases 
the values of $f$ and $\alpha$ satisfy \eqref{eq:f0alpha0u0}
in the region $\mathscr{V}_-$ of the solutions, respectively.
The radius of the surface layer $r_{\rm sl}$ increases and
the ratio $\Delta r/r_{\rm sl}$ decreases as $\lambda$ decreases.
Then, the maximum gravastars could be described by the approximation of
junction with a thin-shell.

\subsection{
Mass of maximum gravastar for various $\lambda$ and $\eta/M_P$}

\begin{figure}[H]
\centering
\includegraphics[width=8cm]{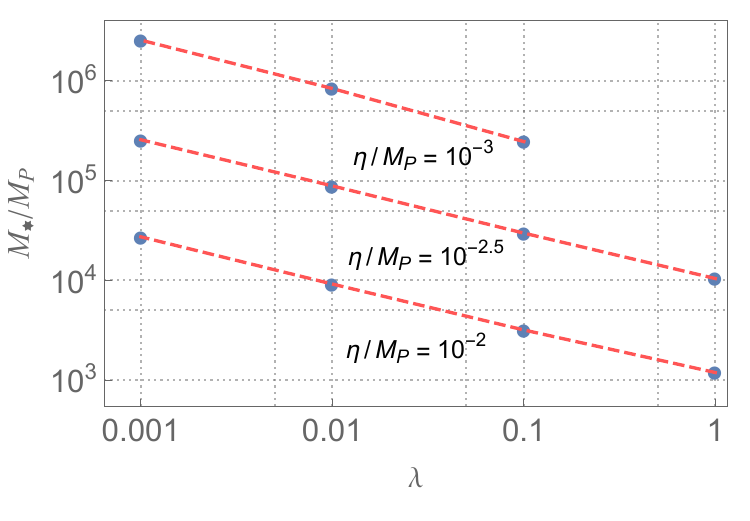}~~~
\includegraphics[width=8cm]{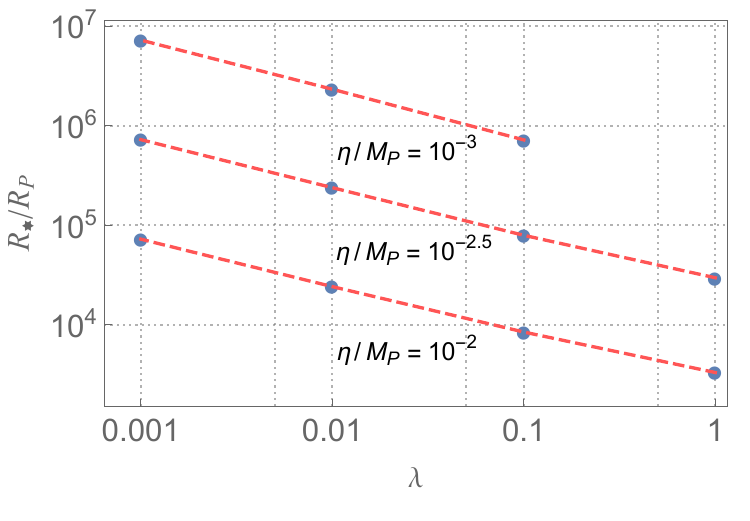}
\caption{
Gravitational mass $M_{\ast}$ and radius $R_{\ast}$ of the maximum gravastars
for various sets of parameters $(\lambda, \eta/M_P)$.
Dashed lines represent the relationships between $M_{\ast}$ and $\lambda$ (left panel), and $R_{\ast}$ and $\lambda$ (right panel) for a fixed value of $\eta/M_P$.
}
\label{fig:Mmax_λ_plot}
\end{figure}

Here, we define the dimensionful radius of the solitonic gravastar as $R:=r_s/\eta$.
In particular the mass and the radius of the maximum gravastar is denoted by $M_{\ast}$ and $R_{\ast}$.
We plot $M_{\ast}$ and $R_{\ast}$ for various sets of $(\lambda,\eta/M_P)$
in Fig.\ref{fig:Mmax_λ_plot}.
We see that $M_{\ast}$ and $R_{\ast}$ obey power laws of $\lambda$ and $\eta/M_P$ as
\begin{align}
  M_{\ast}&\propto\lambda^{-1/2}\left(\frac{\eta}{M_P}\right)^{-2}M_P,
  \label{eq:Mmax_estimate} \\
  R_{\ast}&\propto\lambda^{-1/2}\left(\frac{\eta}{M_P}\right)^{-2}R_P,
  \label{eq:R_estimate}
\end{align}
where $R_P:=G^{1/2}$.
Using the relations \eqref{eq:Mmax_estimate} and \eqref{eq:R_estimate}, we obtain
\begin{align}
  C_{\ast}=\frac{2GM_{\ast}}{R_{\ast}}=const.
\end{align}
From the numerical results, we find that $C_{\ast}\sim 0.7 >2/3$.
Thus, both the ISCO and the photon spere exist in the exterior region of the maximum gravastars regardless of the parameters $(\lambda, \eta/M_P)$ considered in this paper.

\begin{figure}[H]
\centering
\includegraphics[width=8cm]{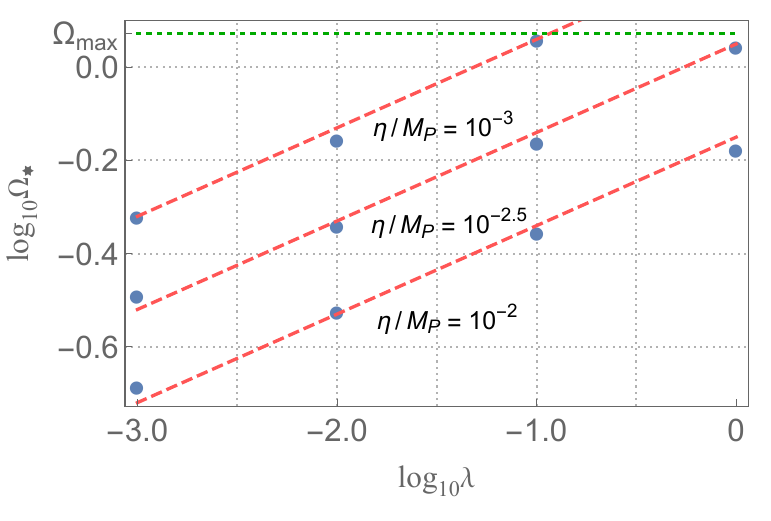}
\caption{
Values of $\Omega_{\ast}$ for various sets of parameters $(\lambda,\eta/M_P)$.
Dot-dashed line represents maximum value of $\Omega$, $\Omega_{max}=\sqrt{\mu}$.
Dashed lines represent the relation of $M_{\ast}$ and $\lambda$ for fixed $\eta/M_P$ by using the formula \eqref{eq:Omega_lambda_eta_relation}.
}
\label{fig:Omega_λ_plot}
\end{figure}

We also plot $\Omega_{\ast}$ for various sets of $(\lambda, \eta/M_P)$
in Fig.\ref{fig:Omega_λ_plot}.
We find an approximate relation between $\Omega_*$ and ($\lambda, \eta/M_P$) as
\begin{align}
  \Omega_*^5 \sim 10^{-5} \frac{\lambda}{(\eta/M_P)^2}
  \label{eq:Omega_lambda_eta_relation}
\end{align}
Since $\Omega_{\ast}$ is limitted by $\Omega_{max}=\sqrt{\mu}\sim 1.183$, we obtain
the upper bound of $\lambda$ as
\begin{align}
  	\lambda \lesssim
	\lambda_{max}(\eta) \sim 10^5\left(\frac{\eta}{M_P}\right)^2.
\label{eq:relation_lambda_eta}
\end{align}
The maximum gravastar
exists for the parameter sets
satisfying \eqref{eq:relation_lambda_eta}.
For the maximum gravastars with $\Omega_{\ast}=\Omega_{max}$, i.e.,
for $\lambda = \lambda_{max}$,
we obtain
\begin{align}
  M_{\ast}\big|_{\Omega_{\ast}=\Omega_{max}}&\sim 10^{-3.5}\left(\frac{\eta}{M_P}\right)^{-3}M_P,
\label{eq:maximumMass_maximumGravastar}
\end{align}
This gives the upper limit of the gravitational mass of solitonic gravastars for given $\eta$.
By using the formula \eqref{eq:Mmax_estimate} and extrapolating the values of $\lambda$ and $\eta$, we summarize the values of $M_{\ast}$ in Table \ref{table:estimete_Mmax}.
We conclude that if $\lambda$ is not extremely small, the mass of a solitonic gravastar is quite light.

\begin{table}[htb]
  \centering
    \begin{tabular}{c|c|c}
        \hline
        Symmetry Breaking Scale & $~~\lambda~~$ & $M_*[\text{kg}]$ \\
        \hline
        \hline
        \multirow{3}{*}{$\eta \sim 10^{16}$ GeV} & $10^{-13}$ & $\mathcal{O}(10^{3})$ \\
        & $10^{-3}$ & $\mathcal{O}(10^{-2})$ \\
        & $\lambda_{max}=10^{-1}$ & $\mathcal{O}(10^{-3})$ \\
        \hline
        \multirow{2}{*}{$\eta \sim 10^{11}$ GeV} & $10^{-23}$ & $\mathcal{O}(10^{18})$ \\
        & $\lambda_{max}=10^{-11}$ & $\mathcal{O}(10^{12})$ \\
        \hline
        \multirow{2}{*}{$\eta \sim 10^{5}$ GeV} & $10^{-30}$ & $\mathcal{O}(10^4M_{\odot})$ \\
        & $\lambda_{max}=10^{-23}$ & $\mathcal{O}(M_{\odot})$ \\
        \hline
    \end{tabular}
    \caption{
        Estimations of the mass of the maximum gravastars, $M_{\ast}$, for various $\lambda$ and $\eta$.
        The upper bound  $\lambda_{max}$ for $\eta$ is given by \eqref{eq:relation_lambda_eta}.
    The symbol $M_\odot$ denotes the solar mass.
    }
    \label{table:estimete_Mmax}
\end{table}

\section{Summary}
We studied a U(1) gauge Higgs model coupled to Einstein gravity.
The model consists of a complex scalar field, a U(1) gauge field, and
a complex Higgs scalar field that causes spontaneous symmetry breaking by the
potential $V(\phi)$ characterized by a coupling constant $\lambda$ and breaking scale $\eta$
in the form
\begin{align}
	V(\phi)=\frac{\lambda}{4}(|\phi|^2-\eta^2)^2.
\end{align}
Numerically, we obtained spherically symmetric nontopological soliton solutions
that are protected by the conserved U(1) charge.
The solutions include a parameter $\Omega$, which denotes the difference
in phase rotation velocities between the two complex scalar fields.

We found solitonic gravastar solutions such that the interior de Sitter spacetime
and the exterior Schwarzschild spacetime are attached by a spherical surface layer
with a finite thickness.
At the surface layer, an electric double layer is produced by the charge of the two complex
scalar fields. The electric field provides the largest component of
the energy-momentum tensor on the surface layer.
By varying the parameter $\Omega$ we found solitonic gravastars
with various masses and compactnesses.
Among them, we identified the maximum gravastar solution with the maximum mass.
The maximum gravastar is compact enough to have both an ISCO and a photon sphere.

We also obtained solutions for various sets of $(\lambda, \eta)$.
Maximum gravastar solutions appear for each set of $(\lambda, \eta)$ in the range
$\lambda \lesssim 10^5(\eta/M_P)^2$ for $\mu=1.4$.
Although we fix the coupling constant at $\mu=1.4$ in this paper, we have confirmed that maximum gravastar solutions exist over a wide range of $\mu$.
Focusing on the maximum gravastars, we derived the relation between the maximum mass, $M_*$,
and the parameters $\lambda$ and $\eta$ in a power-law form.
For $\lambda\sim 0.1$ and $\eta\sim 10^{16}$ GeV, we obtain $M_*\sim 1$g.
This would be an alternative to primordial black holes or would be a seed for their formation.
If we can extrapolate the power-law of $M_*$, the mass could reach an astrophysical scale for much smaller values of $\lambda$ and $\eta$.

There are important works \cite{Visser:2003ge, Nakao:2018knn, Sakai:2014pga, Rosa:2024bqv, Pani:2009ss, Nakao:2022ygj, Uchikata:2016qku, Cardoso:2017cfl} on gravastars
using the thin-shell approximation.
It is a natural question whether a thin-shell approximation exists that can describe solitonic gravastars.
It would be interesting to study the physical phenomena analyzed with the thin-shell approximation using solitonic gravastars in U(1) gauge Higgs models in future work.

\section*{Acknowledgements}

We would like to thank K.-i. Nakao, H. Yoshino, T. Harada, and F. Takahashi for valuable discussions and comments.
This work was partly supported by Osaka Central Advanced Mathematical Institute: MEXT Joint Usage/Research Center on Mathematics and Theoretical Physics No. JPMXP0723833165.
T.O. was supported by JSPS KAKENHI Grant Number 20H05851.

\newpage

\appendix

\section{Energy-momentum tensor}
\label{appendixA}

We show the energy-momentum tensor \eqref{eq:T_munu} for the stationary and spherically
symmetric scalar fields and gauge field in forms of \eqref{eq:matter_variable}, as follows:
\begin{align}
T^t_{~t}/\eta^4=&-\epsilon
\notag \\
=&-\frac{\left(e^2f^2\alpha^2+(e\alpha-\Omega)^2u^2\right)}{\sigma^2(1-2m/r)}-\left(1-\frac{2m}{r}\right)\left(\left(\frac{df}{dr}\right)^2+\left(\frac{du}{dr}\right)^2\right)
\cr
  &-\frac{\lambda}{4}(f^2-1)^2-\mu f^2u^2-\frac{1}{2\sigma^2}\left(\frac{d\alpha}{dr}\right)^2,
 \label{eq:dimensionless_Ttt}
\\
T^r_{~r}/\eta^4=&p_r=p_{\perp}
\notag \\
=&\frac{\left(e^2f^2\alpha^2+(e\alpha-\Omega)^2u^2\right)}{\sigma^2(1-2m/r)}+\left(1-\frac{2m}{r}\right)\left(\left(\frac{df}{dr}\right)^2+\left(\frac{du}{dr}\right)^2\right)
\cr
  &-\frac{\lambda}{4}(f^2-1)^2-\mu f^2u^2-\frac{1}{2\sigma^2}\left(\frac{d\alpha}{dr}\right)^2,
 \label{eq:dimensionless_Trr}
 \\
T^{\theta}_{~\theta}/\eta^4=&p_{\theta}
	=T^{\varphi}_{~\varphi}/\eta^4=p_{\varphi}=p_{\parallel}
\notag \\
=&
\frac{\left(e^2f^2\alpha^2+(e\alpha-\Omega)^2u^2\right)}{\sigma^2(1-2m/r)}-\left(1-\frac{2m}{r}\right)\left(\left(\frac{df}{dr}\right)^2+\left(\frac{du}{dr}\right)^2\right)
\cr
  &-\frac{\lambda}{4}(f^2-1)^2-\mu f^2u^2+\frac{1}{2\sigma^2}\left(\frac{d\alpha}{dr}\right)^2,
 \label{eq:dimensionless_Tthetatheta}
\end{align}
where $\epsilon$ represents an energy density of the fields, $p_r$ and $p_{\theta}$ denote
pressure in the direction of $r$ and $\theta$.
We use dimensionless quantities in these equations by using \eqref{dimensionless}.


\end{document}